\documentclass[draft, preprintnumbers, preprint, nofootinbib]{revtex4}

\usepackage{amsmath, amssymb}
\usepackage[varg]{txfonts} 

\newcommand{\bal}{\begin{equation}\begin{aligned}}
\newcommand{\eal}{\end{aligned}\end{equation}}
\newcommand{\Sp}[1]{S$^#1$}
\newcommand{\ri}{r_\infty}

\begin{document}

\vspace*{5cm}

\bigskip
\title{ \Large
Kaluza-Klein Black Holes with Squashed Horizons
\vspace{1cm}
}

\author{\Large  
	Hideki Ishihara\footnote{E-mail: ishihara@sci.osaka-cu.ac.jp} and
  	Ken Matsuno\footnote{E-mail: matsuno@sci.osaka-cu.ac.jp} }
\address{ \bigskip
  	Department of Mathematics and Physics, Graduate School of Science, \\
	Osaka City University, Sumiyoshi-ku  Osaka  558-8585, Japan
\vspace{2cm}
  }
\preprint{OCU-PHYS 235}
\preprint{AP-GR 27}


\begin{abstract}
We study geometrical structures of charged static black holes 
in the five-dimensional Einstein-Maxwell theory. The black holes 
we study have horizons in the form of squashed $ {\rm S}^3$, 
and their asymptotic structure consists of a twisted ${\rm S}^1$ 
bundle over the four-dimensional flat spacetime at the spatial 
infinity. 
The spacetime we consider is fully five-dimensional in the 
vicinity of the black hole and four-dimensional with a compact 
extra dimension at infinity.
\end{abstract}

\maketitle

\section{Introduction}
Most attempts to construct theories in which gravity is unified with other forces 
are devoted to higher-dimensional spacetimes. 
Non-perturbative gravitational objects in higher-dimensional systems 
might provide insight for a quantum theory of gravity.  
In particular, there are many studies of higher-dimensional black holes, 
starting from pioneering works 
carried out several decades ago\cite{bh_pioneer, bh_old, Myers-Perry, Myers}.

Recently, the idea of large extra dimensions\cite{ADD} has attracted much attention, 
because it suggests the interesting possibility that the unification of 
the electroweak and Planck scales takes place at the TeV scale. 
One of the most interesting phenomena predicted within this scenario is 
the possibility of the formation in accelerators of  
higher-dimensional black holes smaller than the size of extra dimensions\cite{accelerator}. 
Such higher-dimensional black holes would 
reside in a spacetime that is approximately isotropic in the vicinity of the black holes,  
but effectively four-dimensional far from the black holes\cite{Myers}.
We call higher-dimensional black holes with this property 
{\it Kaluza-Klein black holes}.

In this article, we study the geometry of a static charged black hole 
in the five-dimensional Einstein-Maxwell theory. 
The horizons of the black holes have the form of a squashed \Sp3, 
and the spacetime is asymptotically locally flat;  
i.e., it asymptotically approaches 
a twisted \Sp1 bundle over four-dimensional Minkowski spacetime. 
In other words, the spacetime is that of a five-dimensional Kaluza-Klein black hole. 

In the context of the unification of interactions, the Einstein-Maxwell system 
with a Chern-Simons term is a theory of renewed interest 
as the bosonic part of supergravity\cite{Gunaydin}.
There are many works treating black hole solutions 
in this system\cite{susy_BH, GGHPR}. 
Although the Chern-Simons term is absent in our action, 
this has no effect on the solutions discussed below. 
The solutions presented in this article are characterized 
by three parameters: the size of the \Sp1 fiber at infinity, 
the size of the inner horizon, and the size of the outer horizon. 
The singularity of the present static solution is hidden in the horizons, 
which are deformed owing to the non-trivial asymptotic structure. 
This contrasts with the fact that horizons are deformed by the rotation 
of rotating black holes. 
We also discuss some limiting cases of the solutions.

\section{Solutions}
We examine the five-dimensional Einstein-Maxwell theory described 
by the action
\bal
	S=\frac{1}{16\pi G}\int d^5x \sqrt{-g}\left( R 
	-F_{\mu\nu}F^{\mu\nu} \right),
\label{action}
\eal
where $R$ is the scalar curvature, $F=dA$ is the Maxwell field strength 2-form corresponding 
to the gauge potential 1-form $A$, 
and $G$ denotes the five-dimensional Newton's constant.
The equations of motion derived from the action \eqref{action} 
are 
\bal
	d{~}^*F=0 
\quad \mbox{and}\quad 
	R_{\mu\nu}- \frac12R g_{\mu\nu}
	= 2\left(
		F_{\mu\lambda}F_\nu^{~\lambda}
		-\frac14 g_{\mu\nu}F_{\alpha\beta}F^{\alpha\beta}\right).
\label{Einstein_eq}
\eal

Equation 
\eqref{Einstein_eq} admits 
an electrically charged static black hole as a sourceless solution. 
The metric of the solution is 
\bal
 ds^2 = -  f dt^2 +  \frac{k^2}{f} dr^2
       +\frac{r^2}{4}\left[ 
	 k \left\{ (\sigma^1)^2+(\sigma^2)^2 \right\} + (\sigma^3)^2 \right],
\label{metric}
\eal
where $f$ and $k$ are functions of $r$ defined by
\begin{align}
	f(r) = \frac{(r^2-r_+^2)(r^2-r_-^2)}{r^4}, \quad
 	k(r) = \frac{(\ri^2-r_+^2)(\ri^2-r_-^2)}{(\ri^2-r^2)^2},
\label{squashing_fn}
\end{align}
and the gauge potential is
\bal
	A=\pm\frac{\sqrt{3}}{2}\frac{r_+r_-}{r^2} dt.
\label{gauge_pot}
\eal
Here, $r_\pm$ and $\ri$ are constants, and the quantities 
$\sigma^i~ (i=1,2,3)$ satisfy the relation 
\begin{align}
 	d\sigma^{i}=\frac{1}{2}C^i_{jk}\sigma^{j}\wedge \sigma^{k},~\text{with}~ 
 	C^{1}_{23}=C^{2}_{31}=C^{3}_{12}=1~\text{and} ~C^i_{jk}=0 ~\text{in all other cases.}
\end{align}
The static spacetime of the metric \eqref{metric} has the 
isometry group SO(3)$\times$U(1).
Because the metric is apparently singular at $r=\ri$, 
the radial coordinate $r$ should be assumed to move the range 
$ 0 < r <\ri$. 
If we choose the parameters to satisfy $0< r_- \leq r_+ <\ri$,  
the metric has horizons at $r=r_+$, the outer horizon, and 
at $r=r_-$, the inner horizon. 

\section{Shape of horizons}
A time-slice $t=const.$ of the spacetime, which is orthogonal to 
the time-like Killing vector, is foliated by 
three-dimensional surfaces, which are specified by $r=const.$, say $\Sigma_r$.  
Each surface $\Sigma_r$ is regarded as a Hopf bundle, an 
\Sp1 fiber over the \Sp2 base space, with the metric
\begin{align}
	ds^2_{\Sigma_r}&=\frac{r^2}{4}\left[ 
	 k(r) \left\{ (\sigma^1)^2+(\sigma^2)^2 \right\} +(\sigma^3)^2 \right] 
	 =	\frac{r^2}{4}\left[ k(r) d\Omega_{S^2}^2 + \chi^2 \right] ,
\label{Sigma_metric}
\end{align}
where
\begin{gather}
	d\Omega_{S^2}^2=d\theta^2+\sin^2\theta d\phi^2, \quad
	\chi=\sigma^3=d\psi+\cos\theta d\phi. \notag \\                  
	(0 \leq \theta < \pi, ~0 \leq \phi < 2\pi,~ 0 \leq \psi < 4\pi) \label{base_fiber}
\end{gather}
The surface $\Sigma_r$ takes the form of a deformed \Sp3 on which SO(3)$\times$U(1) acts 
as an isometry group. 
The aspect ratio of the \Sp2 base space to the \Sp1 fiber, 
which characterizes the squashing of \Sp3, 
is denoted by the function $k(r)$. 
The degree of squashing increases monotonically 
as $r$ increases towards $\ri$. 
The divergence of $k$ at $\ri$ indicates the collapse of \Sp3 to \Sp2 there.

The shapes of $\Sigma_\pm$, time-slices of the outer and inner horizons, 
are described by the three-dimensional metric \eqref{Sigma_metric} 
with $r=r_\pm$, 
respectively. 
The squashing function $k(r)$ on $\Sigma_\pm$ is
\bal
 	k(r_\pm) =  \frac{\ri^2-r_\mp^2}{\ri^2-r_\pm^2}.
\eal

Because $k(r_+)\geq 1 \geq k(r_-)$,  
the outer horizon is \lq oblate\rq, with \Sp2 larger than \Sp1, 
while the inner one is \lq prolate\rq, with \Sp2 smaller than \Sp1.
In the degenerate case, $r_+=r_-$, the shape of the horizon is 
the round \Sp3. 

\section{Asymptotic structure}
The apparent singularity at $r=\ri$ of the metric \eqref{metric} is the spatial infinity. 
To observe this, we introduce a new radial coordinate $\rho$ as
\bal
	\rho =\rho_0 \frac{r^2}{\ri^2-r^2}, 
\eal
where
\bal
	&\rho_0^2 = k_0 \frac{\ri^2}{4}, \quad
	&k_0=k(0)= f_\infty= f(r_\infty)
	=\frac{(\ri^2-r_+^2)(\ri^2-r_-^2)}{\ri^4}.
\eal
The new coordinate $\rho$ varies from $0$ to $\infty$ 
when $r$ varies from $0$ to $\ri$. 
The metric \eqref{metric} can be rewritten in terms of $\rho$ and 
$T=\sqrt{f_\infty}~t$ as
\bal
 ds^2 = - V dT^2 +\frac{K^2}{V} d\rho^2
		+R^2 d\Omega_{S^2}^2 + W^2 \chi^2, 
\label{metric_far}
\eal
where $V, K, R$ and $W$ are functions of $\rho$ in the form
\bal
	&V=\frac{(\rho-\rho_+)(\rho-\rho_-)}{\rho^2}, \quad
	K^2 = \frac{\rho+\rho_0}{\rho}, \\
	&R^2=\rho^2 K^2 ,\quad
	W^2=\frac{\ri^2}{4}~K^{-2}
		=(\rho_0+\rho_+)(\rho_0+\rho_-)~ K^{-2}.
\label{metric_comp}
\eal
In \eqref{metric_comp} we have used new parameters defined by
\bal
	\rho_\pm = \rho_0 \frac{r_\pm^2}{\ri^2-r_\pm^2}.
\eal

As $\rho\rightarrow\infty$, i.e., $r\rightarrow \ri$, 
the metric \eqref{metric_far} with \eqref{metric_comp} approaches 
\bal
 	ds^2 = - dT^2 + d\rho^2
		+\rho^2 d\Omega_{S^2}^2 + \frac{\ri^2}{4} \chi^2.
\eal
Actually, in the limit of large $\rho$, the leading-order term of 
the Kretschmann scalar is
\bal
	R_{\mu\nu\lambda\sigma}R^{\mu\nu\lambda\sigma} 
		\sim \frac{12 \left[ (\rho_++\rho_-)^2 -\rho_+\rho_- 
		+2(\rho_0+\rho_+)(\rho_0+\rho_-) \right] }{\rho^6} .
\eal
Therefore, the limit $\rho\rightarrow\infty$, and equivalently $r\rightarrow \ri$,  
corresponds to the spatial infinity, and  the spacetime is locally 
asymptotically flat, i.e., 
topologically not a direct product but a twisted \Sp1 fiber bundle
over four-dimensional Minkowski spacetime. 

\section{Physical properties}
First, let us consider the physical properties near 
a black hole. We use the metric form \eqref{metric}. 
In the near region, where $r \ll \ri$, 
the function $k(r)$ is nearly constant, 
and therefore the hyper-area of $\Sigma_r$, i.e., squashed \Sp3, 
is proportional to $r^3$ . 
Thus, the coordinate $r$ plays the role of the hyper-area 
radius of $\Sigma_r$. 
The temporal component $f$ of the metric in coordinates in which $r$ plays such a role  
has the same form as that of the five-dimensional Reissner-Nordstr\"om 
black hole\cite{bh_pioneer}, 
and hence in the region $r \ll \ri$, 
the spacetime we consider is similar to that of a Reissner-Nordstr\"om black hole. 
For the case $r_+ \ll \ri$, there exists an outer region 
satisfying $r_+ < r \ll \ri$, and  
the spacetime would behave as that of a five-dimensional black hole for observers 
in this region. 

Inside the horizon, 
$\Sigma_r$ shrinks to a point with the constant $k_0$ as $r\rightarrow 0$. 
It is obvious that a time-like singularity exists at $r=0$,  
as in the case of the Reissner-Nordstr\"om spacetime. 
The Kretschmann scalar diverges as 
\bal
	R_{\mu\nu\lambda\sigma}R^{\mu\nu\lambda\sigma} 
		\sim 508 r_+^4r_-^4/\left( k_0^8r^{12} \right) 
\eal
in the limit $r \rightarrow 0$. 
The value of $k_0$ denotes the elongation of the central singularity.

Next, we consider the region far from the black hole by using the metric form 
\eqref{metric_far}. 
In the region $\rho_0 \ll \rho$ (equivalently $\ri-r \ll \ri$ 
in the original coordinates), the function $K^2$ is almost constant. 
Hence, the spacetime is effectively four-dimensional for phenomena 
with energies that are lower than the scale of the inverse size of the extra dimension. 
If $\rho_0 \ll \rho_+$, $V$ depends on $\rho$ more strongly than $K^2$. 
In this case, the spacetime would behave as that of 
a four-dimensional Reissner-Nordstr\"om black hole for distant 
observers. 

The size of the outer horizon for a distant observer can be regarded as 
the circumference radius of the \Sp2 base space on the horizon: 
\bal
	R_+=\frac{r_+}{2} \sqrt{\frac{\ri^2-r_-^2}{\ri^2-r_+^2}}.
\eal
When the size of the \Sp1 fiber on the outer horizon, $r_+$, approaches the 
size of the fifth dimension at infinity, $\ri$,  
we could effectively obtain a very large four-dimensional black hole. 

The surface gravity of the outer horizon, $\kappa_+$, is given by
\bal
	\kappa_+ 
		= \frac{\ri^2}{r_+^3}
			\frac{r_+^2-r_-^2}{\ri^2-r_-^2}\sqrt{\frac{\ri^2-r_+^2}{\ri^2-r_-^2}} 
		= \frac{\rho_+-\rho_-}{2\rho_+^2}
			\sqrt{\frac{\rho_+}{\rho_+ +\rho_0}}.
\label{surface_g}
\eal
For the degenerate case, $r_+=r_-$ (equivalently $\rho_+=\rho_-$), 
the surface gravity of 
the black hole is vanishing, as in the usual case.

Using the time-like Killing vector 
$\xi=\partial_t/\sqrt{f_\infty}=\partial_T$, 
which is normalized at the spatial infinity, 
we can define the mass of the black hole as 
\bal
	M = -\frac{3}{32\pi G}\int_\infty dS_{\mu\nu}\nabla^\mu\xi^\nu,
\eal
where the integral is taken over the three-dimensional topological 
sphere at the spatial infinity. 
For the metric form \eqref{metric} with \eqref{squashing_fn}, or 
\eqref{metric_far} with \eqref{metric_comp},
we find  
\bal
	M = \frac{3 \pi}{8 G \sqrt{f_\infty} }\left(r_+^2+r_-^2-\frac{2r_+^2r_-^2}{\ri^2}\right) 
	  = \frac{3 \pi r_\infty}{4 G} ( \rho_+ + \rho_-).
\eal

The electric charge is defined as 
\bal
	Q = \frac{1}{8\pi G}\int_S dS_{\mu\nu}F^{\mu\nu}, 
\eal
where the integral is taken over the three-dimensional topological 
sphere surrounding the black hole. 
For the gauge potential \eqref{gauge_pot}, we have
\bal
	Q = \frac{\sqrt{3} \pi}{2 G} r_+r_- 
	  = \frac{\sqrt{3} \pi \ri}{G} \sqrt{\rho_+\rho_-}.
\eal

\section{Limits}
We consider four kinds of limits of the metric 
parameterized by $r_-, r_+$ and $r_\infty$. 

First, we take the limit $r_\infty \rightarrow \infty$. 
The squashing function behaves as $k\rightarrow 1$ in this limit, and therefore   
the metric \eqref{metric} reduces to that of 
the five-dimensional Reissner-Nordstr\"om black hole 
with the symmetry of round \Sp3, 
i.e., the SO(4) isometry group\cite{Myers-Perry}.  
In this case, the spacetime asymptotically becomes five-dimensional Minkowski spacetime. 
The surface gravity \eqref{surface_g}  
reduces to the well-known form. 

Second, in the limit $r_-\rightarrow 0$, the metric describes 
a neutral black hole with a gravitational radius $r_+$:
\bal
 ds^2 = - \left(1-\frac{r_+^2}{r^2}\right) dt^2 
		+ \left(1-\frac{r_+^2}{r^2}\right)^{-1} k^2 dr^2
        + \frac{r^2}{4}\left[ k d\Omega_{S^2}^2 + \chi^2 \right]. 
\label{neutral_metric}
\eal
When $r_\infty$ is finite, the metric of the black hole 
is deformed by the squashing function $k(r)$ with $r_-=0$. 
The metric in the neutral case has been studied 
in the context of Kaluza-Klein theory\cite{bh_old}. 
If we take the limit  $r_\infty\rightarrow \infty$ in \eqref{neutral_metric}, 
the metric reduces to the five-dimensional 
Schwarzschild black hole with SO(4) isometry, while 
in the limit $r_+\rightarrow 0$,  
the metric reduces to the Gross-Perry-Sorkin monopole\cite{GPS}.

Third, when we set $r_-=r_+$, the metric describes a deformed 
version of the five-dimensional extremal Reissner-Nordstr\"om black hole 
with a degenerate horizon.
In this case, using the new coordinate $\bar r^2 = r^2-r_+^2$ 
and the parameter $\bar r_\infty^2 = r_\infty^2-r_+^2$,  
we can rewrite the metric in the harmonic form
\bal
	ds^2 = -H^{-2} dt^2 + H~ds^2_\text{T-NUT},
\label{extreme}
\eal
where 
\bal
ds^2_\text{T-NUT} = \frac{\bar r _\infty ^8}{(\bar r _\infty ^2-\bar r ^2)^4}d\bar r ^2
	+\frac{\bar r ^2}{4}\left[
		\frac{\bar r _\infty ^4}{(\bar r _\infty ^2-\bar r ^2)^2}d\Omega_{S^2}^2
		+\chi^2\right] \\
\eal
is the metric of the Euclidean self-dual Taub-NUT space in four-dimensions, 
and we have 
\bal
	H=1+\frac{r_+^2}{\bar r ^2}. 
\eal
In the case $r_+ = r_-$ (equivalently $M=\frac{\sqrt{3}}{2}Q$), 
the metric is a special case of the supersymmetric solution to 
 five-dimensional supergravity\cite{GGHPR, Reduced_BH}.

Finally, we consider the limit $r_+, r_- \rightarrow r_\infty$, with  
$\rho_\pm$ finite. It is convenient to use the metric 
\eqref{metric_far}.
Because    
$	\rho_0 \rightarrow 0,~ 
	K^2 \rightarrow 1 
$
in this limit, the metric becomes
\bal
	ds^2 = -\frac{(\rho-\rho_+)(\rho-\rho_-)}{\rho^2} dT^2 
			+ \frac{\rho^2}{(\rho-\rho_+)(\rho-\rho_-)}d\rho^2
	 	+ \rho^2 d\Omega_{S^2}^2 	
			+ \rho_+\rho_- \chi^2. 
\label{const_ext}
\eal
This metric describes a four-dimensional Reissner-Nordstr\"om black hole 
with a twisted \Sp1 bundle, 
where the size of the \Sp1 fiber takes the constant value 
$\sqrt{\rho_+\rho_-}=r_\infty/2$. 
The surface gravity \eqref{surface_g} reduces to that of 
a four-dimensional black hole in this limit. 
Furthermore, if we take the limit $\rho_- \rightarrow 0$ 
in the metric \eqref{const_ext}, after redefining the angular coordinate $\psi$
in \eqref{base_fiber} as $\psi'=\sqrt{\rho_+\rho_-}\psi, $
the metric \eqref{const_ext} approaches 
the direct product metric of the four-dimensional Schwarzschild black hole 
and \Sp1, 
\bal
	ds^2 = - \left(1-\frac{\rho_+}{\rho}\right) dT^2 
		   + \left(1-\frac{\rho_+}{\rho}\right)^{-1} d\rho^2
	 	   + \rho^2 d\Omega_{S^2}^2 + d\psi'^2, 
\label{product}
\eal
whose universal covering space describes a black string.  

It would be interesting to clarify the stability 
and phase structure of five-dimensional gravitational 
objects\cite{stability} by using the family of solutions presented 
in this article. 
It is also important to investigate the thermodynamics 
of black holes including parameters representing 
the sizes of the extra dimensions. 


\section*{Acknowledgements}
We thank K. Nakao, Y. Yasui and D. Ida for useful discussions. 
This work is supported by a Grant-in-Aid
for Scientific Research (No.14540275).




\begin{thebibliography}{99}

\bibitem{bh_pioneer} 
F. R. Tangherlini, Nuov. Cim. {\bf 27}, 636 (1963).

\bibitem{bh_old}
P. Dobiasch and D. Maison, Gen. Rel. Grav. {\bf 14}, 231 (1982); \\
G. W. Gibbons and D. L. Wiltshire, Ann. Phys. {\bf 167}, 201 (1986).

\bibitem{Myers-Perry}
R.~C.~Myers and M.~J.~Perry, Ann. Phys. {\bf 172}, 304 (1986).

\bibitem{Myers}
R.~C.~Myers, Phys. Rev. D {\bf 35}, 455 (1987).

\bibitem{ADD}
N.~Arkani-Hamed, S.~Dimopoulos and G.~R.~Dvali, Phys.\ Lett.\ B {\bf 429}, 263 (1998); \\
I. Antoniadis, N. Arkani-Hamed, S. Dimopoulos and G. R. Dvali, Phys. Lett. B {\bf 436}, 257 (1998). 

\bibitem{accelerator}
T.~Banks and W.~Fischler, hep-th/9906038; \\
S. B. Giddings and S. Thomas, Phys. Rev. D {\bf 65}, 056010 (2002).

\bibitem{Gunaydin}
M.~Gunaydin, G.~Sierra and P.~K.~Townsend, Nucl. Phys. B {\bf 253}, 573 (1985).

\bibitem{susy_BH} 
M.~Cvetic and D.~Youm, Nucl. Phys. B {\bf 476}, 118 (1996); \\
%
D.~Klemm and W.~A.~Sabra, JHEP {\bf 02}, 031 (2001); \\
%
M.~Cvetic, H.~Lu and C.~N.~Pope, Phys. Lett. B {\bf 598}, 273 (2004);\\
%
J.~B.~Gutowski and H.~S.~Reall, JHEP {\bf 04}, 048 (2004).
%
\bibitem{GGHPR}
J. P. Gauntlett, J. B. Gutowski, C. M. Hull, S. Pakis and H. S. Reall, 
Class. Quant. Grav. {\bf 20} 4587, (2003).
%

\bibitem{GPS} 
D. J. Gross and M. J. Perry, Nucl. Phys. B {\bf 226}, 29 (1983); \\
R. D. Sorkin, Phys. Rev. Lett. {\bf 51}, 87 (1983).  

\bibitem{Reduced_BH} 
H. Elvang, R. Emparan, D. Mateos and H. S. Reall, JHEP {\bf 08}, 042 (2005);\\
D. Gaiotto, A. Strominger and X. Yin, JHEP {\bf 02}, 024 (2006).

\bibitem{stability}
R. Gregory and R. Laflamme, Phys. Rev. Lett. {\bf 70}, 2837 (1993);\\
T. Wiseman, Class. Quantum Grav. {\bf 20}, 1137 (2003);\\
B. Kol, JHEP {\bf 10}, 049 (2005).


\end{thebibliography}
\end{document}